%% file: eprint.tex
\documentclass[12pt]{article}
\usepackage[utf8]{inputenc}
\usepackage{graphicx}

\newcommand\pubnumber{TTK-22-43, P3H-22-120}
\newcommand\pubdate{\today}

\def\institute{Institute for Theoretical Particle Physics and Cosmology RWTH-Aachen University, D-52056 Aachen,
Germany}
\def\support{\footnote{The work of Jasmina Nasufi was supported by the Deutsche Forschungsgemeinschaft (DFG) under grant 396021762 - TRR 257: P3H - Particle Physics Phenomenology after the Higgs Discovery and under grant 400140256 - GRK 2497: The physics of the heaviest particles at the Large Hadron Collider.}}

\def\Title#1{\begin{center} {\Large #1 } \end{center}}
\def\Author#1{\begin{center}{ \sc #1} \end{center}}
\def\Address#1{\begin{center}{ \it #1} \end{center}}

\newcommand\pubblock{\rightline{\begin{tabular}{l} \pubnumber\\
         \pubdate  \end{tabular}}}
\newenvironment{Abstract}{\begin{quotation}  }{\end{quotation}}
\newenvironment{Presented}{\begin{quotation} \begin{center} 
             PRESENTED AT\end{center}\bigskip 
      \begin{center}\begin{large}}{\end{large}\end{center} \end{quotation}}

\input econfmacros.tex

\begin{document}
\begin{titlepage}
\pubblock

\vfill
\Title{$t\overline{t}Z$ in the 4$\ell$ channel at NLO in QCD}
\vfill
\Author{ Jasmina Nasufi\support}
\Address{\institute}
\vfill
\begin{Abstract}
\noindent NLO QCD corrections to the process $pp \rightarrow e^+\nu_e \mu^- \bar{\nu}_{\mu} \tau^+\tau^- b \bar{b}+X$ are presented with the full off-shell effects included. The calculation includes all resonant and non-resonant Feynman diagrams, photon and $Z$-gauge boson contributions, as well as interference effects, all incorporated at the matrix element level. Furthermore all heavy intermediate particles are described via Breit-Wigner propagators. Theoretical uncertainties related to the scale variation and choice of PDF sets were investigated at the integrated and differential level. The modelling is studied by a direct comparison of the full off-shell process to the NWA, where tops, $Z$- and $W$-gauge bosons are on-shell. Moreover, motivated by experimental cuts, we also investigate the impact of imposing a window cut around the mass of the $Z$-gauge boson on the full off-shell predictions. 
\end{Abstract}
\vfill
\begin{Presented}
$15^\mathrm{th}$ International Workshop on Top Quark Physics\\
Durham, UK, 4--9 September, 2022
\end{Presented}
\vfill
\end{titlepage}
\def\thefootnote{\fnsymbol{footnote}}
\setcounter{footnote}{0}

\section{Introduction}

The associated production of a top pair and a $Z$-gauge boson is a very important process to study at the LHC. It features some of the heaviest particles known to-date and its signature can therefore be modified by various new physics scenarios (e.g. ref. \cite{Greiner:2014qna,Baur:2004uw}). Furthermore, $t\overline{t}Z$ comprises an important background to the SM process $t\overline{t}H$ (e.g. ref. \cite{ATLAS:2017ztq}) and other BSM processes (e.g. ref. \cite{ATLAS:2016dlg}) in the multi-lepton channels. It has been measured at the LHC by both ATLAS and CMS (e.g. ref \cite{CMS:2019too,ATLAS:2021fzm}). In this work, we focus on $pp \rightarrow e^+\nu_e \mu^- \bar{\nu}_{\mu} \tau^+\tau^- b \bar{b}+X$ at NLO in QCD with all full off-shell effects included. All leptons and $b$-jets are treated as massless. The default fiducial phase space region is defined by applying transverse momentum, rapidity and $\Delta R$ cuts on all final states. For details, please see ref. \cite{Bevilacqua:2022nrm}.

\section{Results}
\begin{figure}[b]
\centering
\includegraphics[scale=0.4]{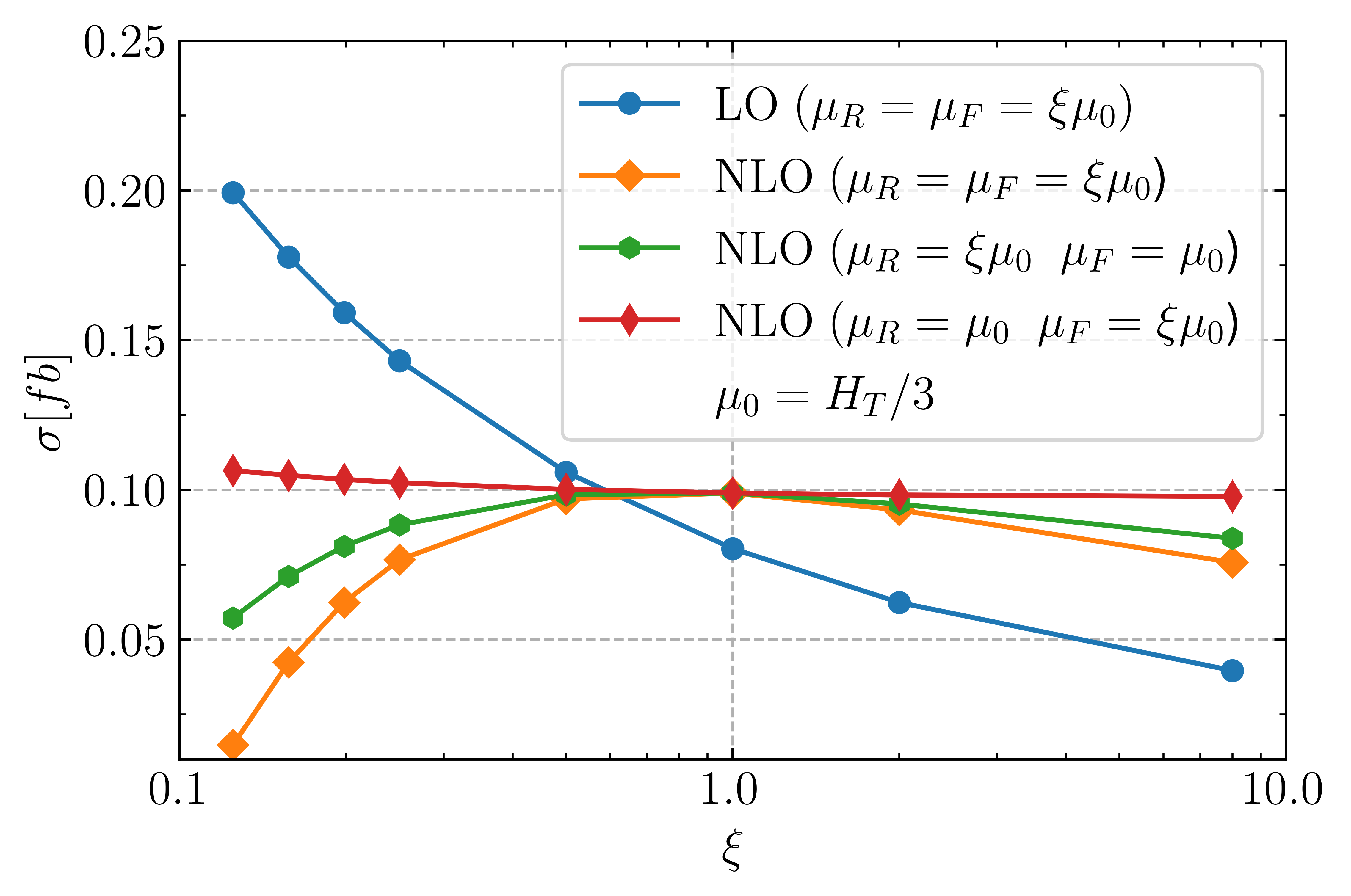}
\includegraphics[scale=0.15]{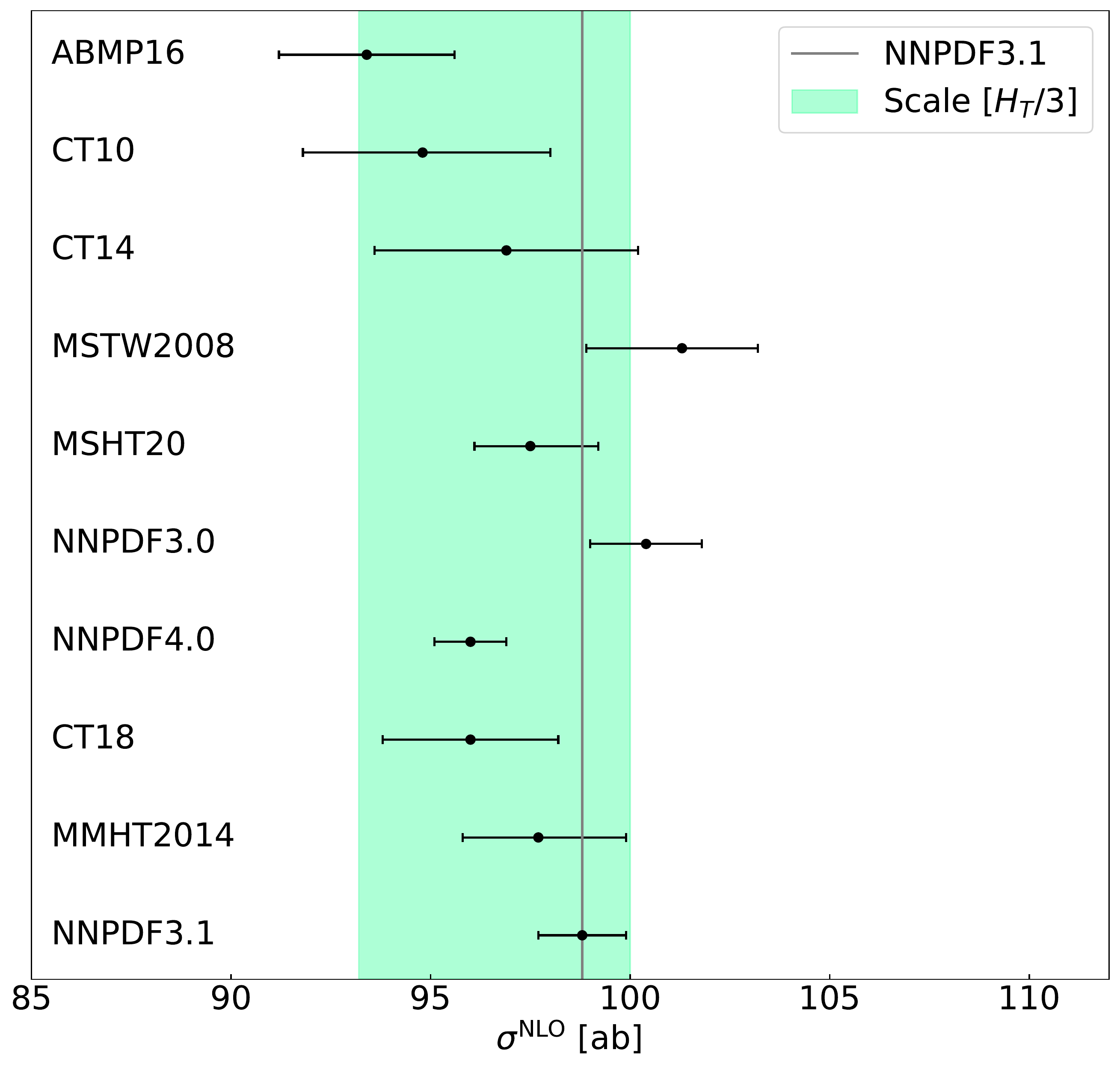}
\caption{(Left) Scale dependance plot for the dynamic scale. (Right) Various PDF sets compared against each other and the scale uncertainty.}
\label{fig1}
\end{figure}
\begin{figure}[t]
\centering
\includegraphics[scale=0.25]{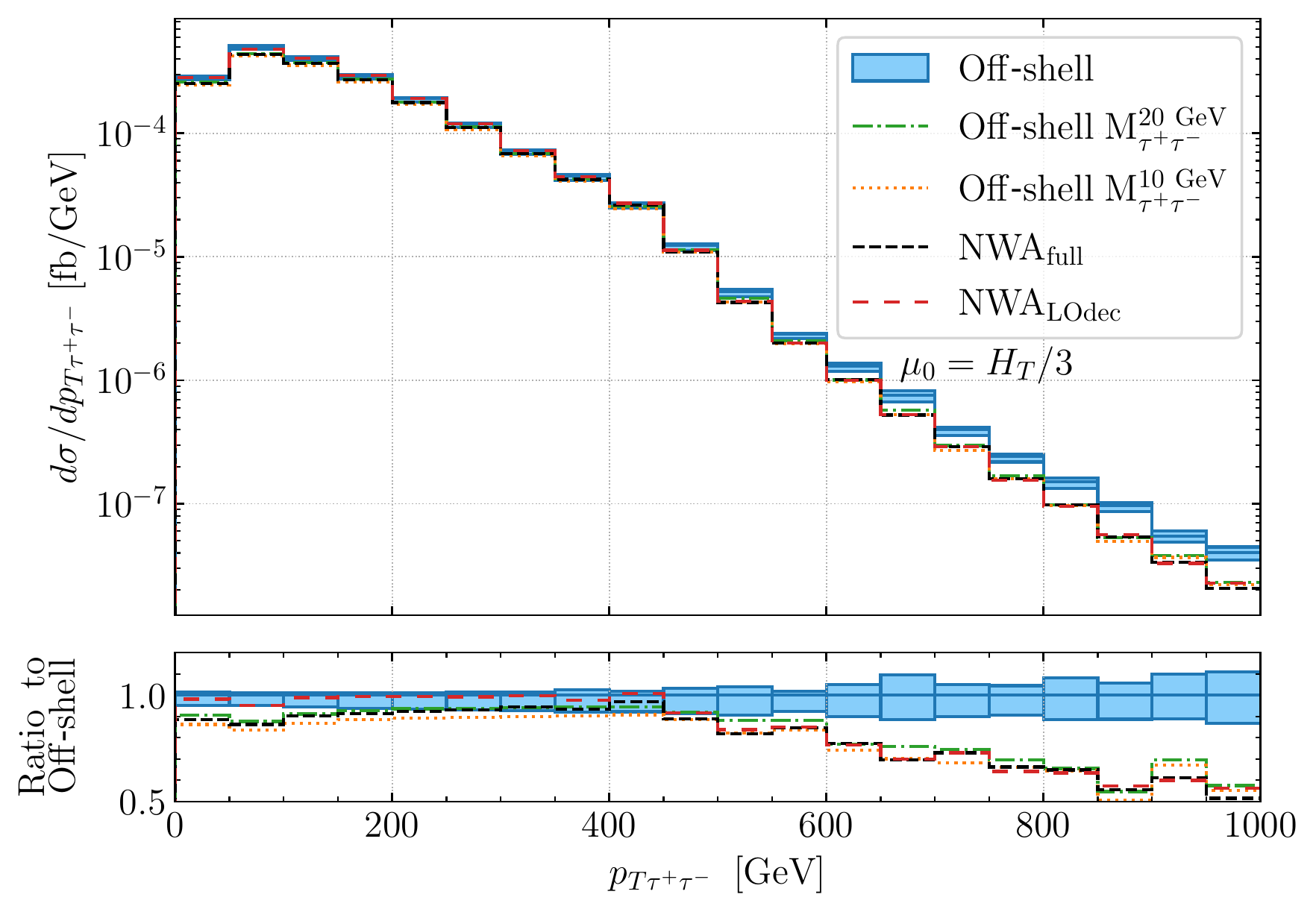}
\includegraphics[scale=0.25]{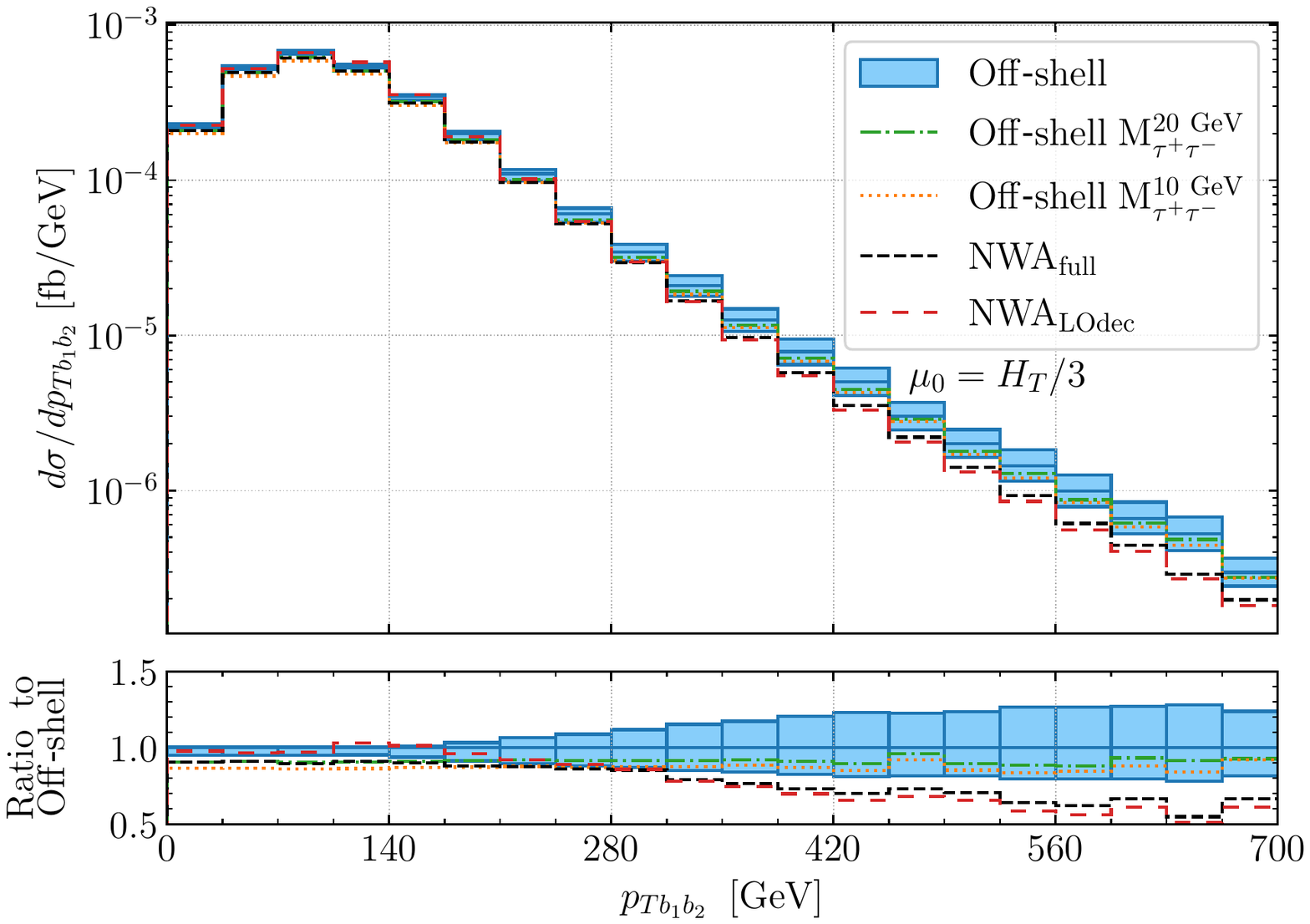}
\includegraphics[scale=0.25]{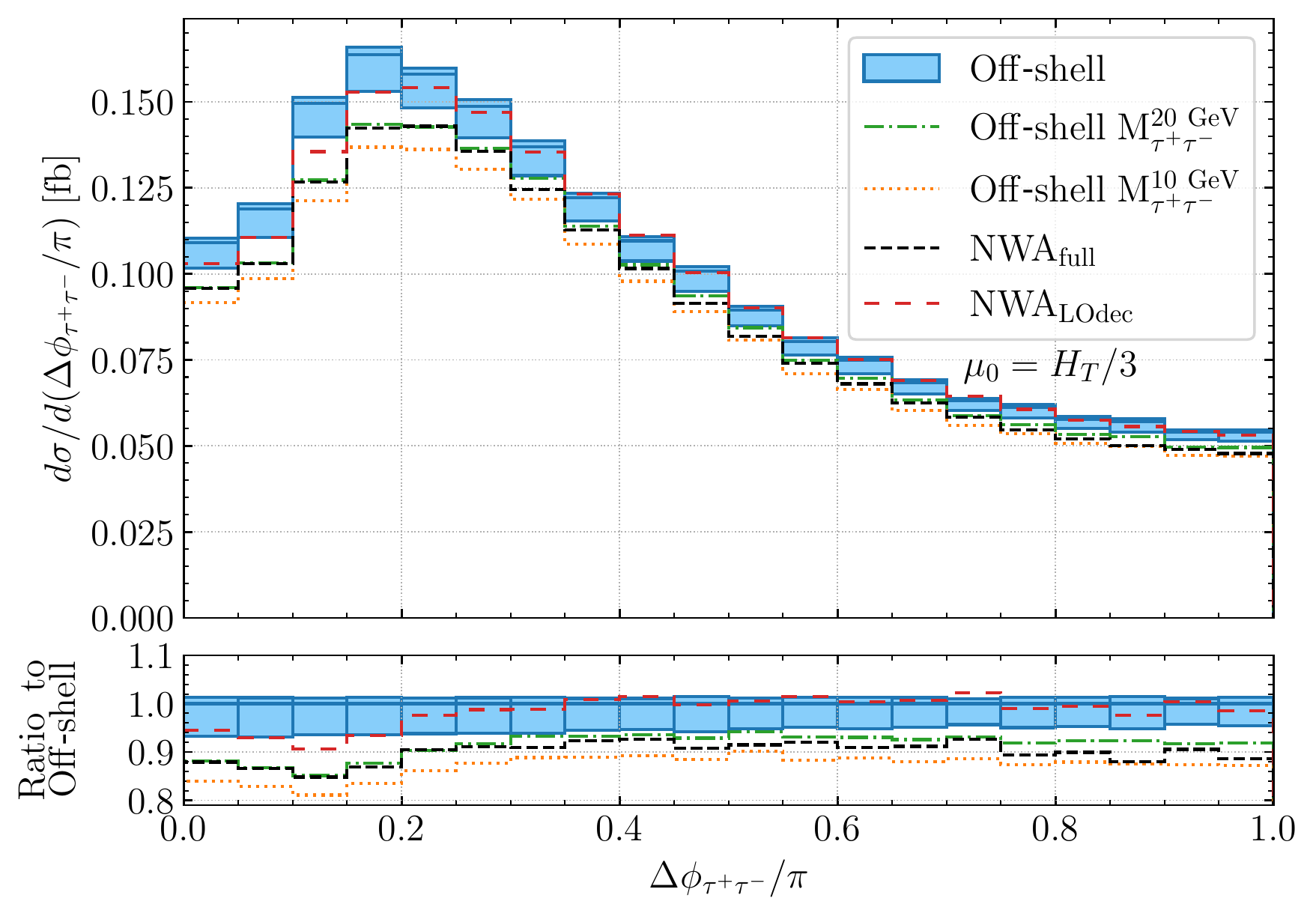}
\caption{Differential fiducial cross sections for $t\bar{t}Z$ with various modelling approaches.}
\label{fig2}
\end{figure}
The integrated fiducial cross sections at LO and NLO in QCD, with a dynamic scale setting and the NNPDF3.1 PDF set are:
\begin{eqnarray}
\sigma^{\rm LO}_{\rm full~off-shell} = 80.32^{+32\%}_{-22\%}~{\rm ab} \hspace*{2cm}
\sigma^{\rm NLO}_{\rm full~off-shell} = 98.88^{+1\%}_{-6\%}~{\rm ab}
\label{eqn1}
\end{eqnarray}
Thus the integrated cross section is increased by $+23\%$ once NLO QCD corrections are applied. The size of the scale uncertainty is reduced to $6\%$ maximally at NLO. On the left of fig. \ref{fig1} the integrated cross section is computed for different re-scalings of the renormalization $\mu_R$ and factorization $\mu_F$ scales by a factor $\xi$, with respect to the central dynamic scale choice. We note that the variation in the integrated cross section is driven primarily by $\mu_R$. Yet another systematic source of theory uncertainties are intrinsic PDF uncertainties. For the NLO cross section in eq. (\ref{eqn1}) the PDF uncertainty is about $\pm 1\%$. Another way of evaluating PDF uncertainties is to compare the predictions computed with different central PDF sets. This is shown on the right of fig. \ref{fig1}, where we note that for most PDF sets the biggest difference to the default NNPDF3.1 PDF set as well as their corresponding intrinsic PDF uncertainties are around $3\%$ or smaller. The difference to NNPDF3.1 is larger only for the ABMP16 PDF set, where it amounts to about $5\%$. In all cases, scale variation  dominates the theoretical uncertainties.\\
Furthermore, we also investigate the modelling in the presence of the full off-shell effects by comparing to the full NWA$_{\rm full}$ and the NWA with LO decays NWA$_{\rm LOdec}$. In the NWA, $\tau^+\tau^-$ can only originate from an on-shell $Z$-gauge boson, where the latter is emitted in the production stage. On the other hand, full off-shell predictions allow for off-mass-shell and $\gamma^* \rightarrow\tau^+\tau^-$ contributions. This leads to sizable full off-shell effects of about $11\%$ at the integrated level. To investigate them further, we impose an additional window cut $|M_{\tau^+\tau^-}-m_Z|<X$, that aims to remove photon induced contributions. By varying $X$ between $25$ GeV and $10$ GeV full off-shell effects are reduced below $3\%$. \\
The impact of this cut can be visualized even better at the differential level in fig. \ref{fig2}. For $Z$-boson related observables such as $p_{T\tau^+\tau^-}$, imposing this cut with $X=10,20$ GeV to the full off-shell predictions leads to better agreement with the  NWA$_{\rm full}$ even in the high $p_T$ regions, where full off-shell effects are sizable. On the other hand, for top-quark related observables, such as $p_{Tb_1b_2}$, the window cut does not remove the single-resonant contributions, visible at the end of the plotted $p_T$ spectrum. Due to the photon induced contributions, full off-shell effects are also visible for angular observables. 

\section{Conclusion}
In conclusion, NLO QCD corrections to $t\overline{t}Z$ in the $4\ell$ decay channel are large ($23\%$) and they reduce the dominant scale uncertainties to at most $6\%$. Full off-shell effects for this process are sizable ($11\%$) due to the photon contribution and interference effects. We find that by imposing a window cut with  $X=10,20$ GeV one can effectively remove the full off-shell effects related to the $Z$-gauge boson.

\clearpage
\bibliography{References} 
\bibliographystyle{JHEP}

\end{document}

%% file: econfmacros.tex
\def\beq{\begin{equation}}
\def\eeq#1{\label{#1}\end{equation}}
\def\eeqn{\end{equation}}

\def\beqa{\begin{eqnarray}}
\def\eeqa#1{\label{#1}\end{eqnarray}}
\def\eeqan{\end{eqnarray}}

\let\bar=\overbar

\def\Dslash{\not{\hbox{\kern-4pt $D$}}}
\def\dslash{\not{\hbox{\kern-2pt $\del$}}}

\def\msb{{\bar{\ssstyle M \kern -1pt S}}}

